\documentclass[particles,article,accept,moreauthors,pdftex,10pt,a4paper]{mdpi} 

\firstpage{1} 
\articlenumber{11}
\doinum{10.3390/particles1010011}
\pubvolume{1}
\pubyear{2018}
\copyrightyear{2018}
\history{Received: 26 April 2018; Accepted: 19 June 2018; Published: 21 June 2018}

\pdfoutput=1

 \theoremstyle{mdpi}
 \newcounter{thm}
 \newcounter{ex}
 \newcounter{re}
 \theoremstyle{mdpidefinition}
\usepackage[T1]{fontenc}
\usepackage{graphicx,mathrsfs}
\usepackage{bm}
\usepackage{amssymb}
\usepackage{amsmath,stmaryrd}
\usepackage{amsfonts}
\usepackage{dcolumn}
\usepackage{natbib}
\usepackage{color}
\usepackage{calc}
\usepackage{longtable}
\usepackage{pdflscape}
\usepackage[normalem]{ulem}
\usepackage{soul}

\newcommand{\be}{\begin{equation}}
\newcommand{\ee}{\end{equation}}
\newcommand{\bea}{\begin{eqnarray}}
\newcommand{\eea}{\end{eqnarray}}
\newcommand{\Tr}{{\rm Tr}}

\newcommand{\raiseentry}[1]{\smash{\raise 0.7 em \hbox{#1}}}

\newcommand{\lowentry}[1]{\smash{\lower 1.5 ex \hbox{#1}}}

\def\prd{Phys. Rev. D.}

\usepackage{booktabs} 
\usepackage{multirow}
\usepackage{soul} 
\usepackage{microtype}

\Title{
Relativistic Dissipative Fluid Dynamics from the
Non-Equilibrium Statistical Operator
         }

\Author{
   {Arus Harutyunyan}  $^{1}$*\orcidA{}, 
   Armen Sedrakian    $^{2}$\orcidB{} and
   Dirk H.\ Rischke      $^{1,3}$
            }

\AuthorNames{Arus Harutyunyan, Armen Sedrakian, and Dirk H.\ Rischke}

\address{\noindent
$^{1}$ \quad \textls[-25]{Institute for Theoretical Physics, Goethe
  University, Max-von-Laue-Stra\ss e, 1,~60438~Frankfurt~am~Main,~Germany; arus@th.physik.uni-frankfurt.de, drischke@th.physik.uni-frankfurt.de} 
$^{2}$ \quad  Frankfurt Institute for Advanced Studies,
Ruth-Moufang-Stra\ss e, 1, 60438 Frankfurt am Main,
  Germany; sedrakian@fias.uni-frankfurt.de 
$^{3}$ \quad  Department of
Modern Physics, University of Science and Technology of China, Hefei 230026, China
               }
 \corres{Correspondence: arus@th.physik.uni-frankfurt.de}

               \abstract{ We present a new derivation of 
                 second-order relativistic dissipative {fluid} dynamics
                 for quantum systems using Zubarev's formalism {for} the
                 non-equilibrium statistical operator.  In particular,
                 we discuss the {shear-stress tensor to second order in gradients}
                 and argue that the relaxation terms for
                 the dissipative quantities arise from {memory
                 effects contained in the statistical operator}.  We also
                 identify new transport coefficients which
                 describe the relaxation of dissipative processes to
                 second order and express them in terms of
                 equilibrium correlation functions, thus establishing
                 Kubo-type formulae for the second-order transport
                 coefficients.  }


\begin{document}
\section{Introduction}
\label{sec:intro}

Fluid dynamics is a powerful tool to describe low-frequency and
long-wavelength phenomena in statistical
systems~\cite{Landau:Hydro}. It finds numerous applications in
astrophysics, cosmology, heavy-ion physics, and other areas. In
particular, it has been successfully applied to describe {the
  collective behavior of hot and dense strongly interacting matter}
created in heavy-ion collision experiments at Relativistic Heavy Ion
Collider (RHIC) and Large Hadron Collider (LHC).  {In these
  experiments, a new state of matter, the~{quark-gluon} plasma
  (QGP), was discovered, which~behaves almost like a perfect fluid.}

There are two main {approaches} which can be used to derive {the
  equations of motion of fluid dynamics and the pertaining} transport
coefficients from the underlying microscopic theory. For~weakly
interacting systems, one commonly relies on kinetic theory based on the
Boltzmann equation for the quasi-particle distribution function
\cite{2014JPhG...41l4004D,Denicol2012,2017arXiv170702282F,2010IJMPE..19....1R}.
For strongly interacting quantum systems, where~the quasi-particle
picture breaks down and/or the quantum nature of the fields itself is
important, kinetic~theory is no longer applicable, and a full
quantum-statistical approach based on the Liouville equation for the
non-equilibrium statistical operator is required.

In this work, we adopt the method of the non-equilibrium statistical
operator
{(NESO)}~\cite{zubarev1974nonequilibrium,zubarev1997statistical} to
obtain the relativistic {fluid-dynamical} equations of motion for
strongly correlated matter, such as the QGP, in the non-perturbative
regime. The method was applied to quantum
fields~\cite{1984AnPhy.154..229H} and has been since extended to treat
systems in strong magnetic fields~\cite{2011AnPhy.326.3075H}.  It is
based on a generalization of the Gibbs canonical ensemble to
non-equilibrium states, i.e., the~statistical operator is promoted to
a non-local functional of the thermodynamic parameters and their
{space-time} derivatives.  Assuming that the thermodynamic
parameters are sufficiently smooth over the correlation lengths
characterizing the system, the~statistical operator is expanded into a
series in gradients of these parameters to the desired order. The
{fluid-dynamical} equations for the dissipative fluxes emerge then
after statistically averaging the relevant quantum operators.  An
advantage of the {NESO method} is that the transport coefficients of
the system are automatically obtained in the form of Kubo-type
relations, i.e., they are related to certain correlation functions of
the underlying field theory in the strong-coupling regime.  There
exist a number of formulations of relativistic fluid dynamics in terms
of near-equilibrium quantities which are related to the NESO method
employed by us; for recent work, see
References~\cite{Hayata2015,Becattini2015,Hongo2017}.

This contribution provides a concise presentation of our recent work
on the derivation of second-order {dissipative} fluid dynamics via the
NESO method~\cite{Hydro_preprint, ArusPhD}.~As~{is} well known,
{relativistic~fluid dynamics} describes the state of a fluid in terms
of its energy-momentum tensor and currents of conserved charges, which
in the relevant low-frequency and long-wavelength limit can be
expanded around their equilibrium values.  The zeroth-order expansion
corresponds to ideal (non-dissipative) {fluid dynamics}. At first
order, dissipative relativistic {fluid dynamics} emerges from a
truncation that keeps the {terms of linear order in
  gradients}~\cite{1940PhRv...58..919E,Landau:Hydro}. Second-order
relativistic theories have also been
constructed~\cite{1976AnPhy.100..310I,1979AnPhy.118..341I} to avoid
the acausality of the first-order theory and the resulting numerical
instabilities.  In second-order theories, the dissipative fluxes
satisfy relaxation equations, which~describe the process of their
relaxation towards their Navier--Stokes values at asymptotically large
times. While~the general structure of second-order fluid dynamics is
known, {different} results have been obtained for the coefficients
entering these equations (see, {e.g.,}
References~\cite{2008JHEP...04..100B,MooreSohrabi2012}). The~various
versions of second-order fluid dynamics and the pertaining relaxation
equations are reviewed and compared to each other, {e.g.,}~in the review
articles~\cite{2010IJMPE..19....1R,2014JPhG...41l4004D,2017arXiv170702282F},
to which we refer the reader for more detailed~expositions.

This work is structured as follows. Section~\ref{sec:Zub_formalism}
gives a brief summary of Zubarev's formalism {for} the
NESO~\cite{zubarev1974nonequilibrium,zubarev1997statistical}. Section~\ref{sec:1_order}
recapitulates Navier--Stokes theory and the Kubo formulae for the
first-order transport coefficients. The second-order transport
equations are discussed in Section~\ref{sec:2_order} and a summary is
given in Section~\ref{sec:summary}.  We work in flat {space-time} 
described by the metric {tensor} $g^{\mu\nu}={\rm diag}(+,-,-,-)$.

\section{Non-Equilibrium Statistical Operator and Correlation Functions}
\label{sec:Zub_formalism}

The {fluid-dynamical} state of a relativistic quantum system is described
by the operators of the energy-momentum tensor $\hat{T}^{\mu\nu}(x)$
and the conserved particle current $\hat{N}^{\mu}(x)$. For~example,
in the case of Dirac fermions, the~particle current is given 
by $\hat{N}^\mu=\hat{\bar{\psi}}\gamma^\mu\hat{\psi}$, where~{$\hat{\psi}$ is the Dirac field operator}, 
and $\gamma^\mu$ are the Dirac matrices. The equations of
relativistic {fluid dynamics} consist of the covariant conservation laws
for these quantities
\be\label{eq:cons_laws}
\partial_{\mu} \hat{T}^{\mu\nu}(x) =0,\qquad
\partial_{\mu} \hat{N}^{\mu}(x) =0.
\ee

Here, we assume that the fluid {consists of} only one particle species. The
generalization to the case of {several} conserved species is
straightforward and is given elsewhere~\cite{Hydro_preprint}.

{In general, the~fluid-dynamical description is applicable, if the actual state of a given system does not
deviate too much from} local
thermodynamic equilibrium. This allows one to introduce {a fictitious local-equilibrium reference state,
characterized by space-time dependent}
thermodynamic parameters, such as temperature
$T(x)\equiv \beta^{-1}(x)$, chemical potential $\mu(x)$, and fluid
4-velocity $u^{\nu}(x)$. {The deviation of the actual state from this fictitious reference state
is then taken to be proportional to gradients of these fields. The assumption that
the deviation from local equilibrium is small is 
equivalent to assuming that these fields} are slowly varying functions of the space-time
coordinates $x\equiv (\bm x,t)$.  Note that, in this context, ``slowly'' means that the characteristic
{\it {macroscopic}} scales over which the {fluid-dynamical} quantities change
in space and time should be much larger than the characteristic
{\it {microscopic}} scales of the system, {{e.g.,} for quasi-particles the mean free path between collisions}.
In terms of {the thermodynamic parameters defined above}, we
define new auxiliary functions
\be\label{eq:beta_nu_alpha}
\beta^\nu(x) = \beta(x) u^{\nu}(x),\qquad
\alpha (x)=\beta(x)\mu (x).
\ee

Now, consider the NESO given by~{Huang et al.} \cite{2011AnPhy.326.3075H}:
\be\label{eq:stat_op_full}
\hat{\rho}(t) = Q^{-1}e^{-\hat{A}+\hat{B}},
\qquad Q=\Tr e^{-\hat{A}+\hat{B}},
\ee
where
\bea\label{eq:A_op}
\hat{A}(t)&=&\int d^3x \Big[\beta^\nu(x)
\hat{T}_{0\nu}(x)-\alpha(x) \hat{N}^0(x)\Big],\\
\label{eq:B_op}
\hat{B}(t)&=& \int d^3x_1 \int_{-\infty}^tdt_1 e^{\varepsilon(t_1-t)} \hat{C}(x_1),\\
\label{eq:C_op}
\hat{C}(x)&=&\hat{T}_{\mu\nu}(x)
\partial^{\mu}\beta^\nu(x)-
\hat{N}^\mu(x)\partial_\mu\alpha(x),
\eea
with $\varepsilon\to +0$ taken after the thermodynamic limit. The NESO
satisfies the quantum Liouville equation with an infinitesimal source
term $\varepsilon$, which~for positive  values selects the retarded
solution~\cite{zubarev1974nonequilibrium, zubarev1997statistical}.
The~operators $\hat{A}(t)$ and $\hat{B}(t)$ correspond to the
equilibrium and non-equilibrium parts of the statistical operator,
where the operator $\hat{C}({x})$ stands for the thermodynamic
``force'' as it involves the gradients of the thermodynamic
variables, i.e., temperature, chemical potential, and fluid
4-velocity. We also define the {\it {local-equilibrium statistical
  operator}} as
\be\label{eq:stat_op_local}
\hat{\rho}_l(t) = Q^{-1}_le^{-\hat{A}},
\qquad Q_l=\Tr e^{-\hat{A}},
\ee
which is the analog of the Gibbs distribution involving local
thermodynamic parameters.

Before proceeding, we remark that the thermodynamic variables are
well-defined quantities only in an equilibrium state,  {but not for a non-equilibrium state.
The reason  they appear at all in our discussion is the introduction of a fictitious
local-equilibrium state, from which the actual non-equilibrium state should not deviate too much.
The freedom in choosing this fictitious state can be exploited to determine 
the parameters $\alpha(x)$, $\beta(x)$, and the fluid 4-velocity $u^\nu(x)$ characterizing this state}. 
For this purpose, we first
define the operators of the energy and particle densities via
$\hat{\epsilon}{(x)}=u_\mu {(x)} u_\nu {(x)} \hat{T}^{\mu\nu} {(x)}$ and
$\hat{n} {(x)}=u_\mu {(x)} \hat{N}^\mu {(x)}$. These simply imply that $\hat{\epsilon}{(x)}$
and $\hat{n}{(x)}$ are the time-like eigenvalues of the energy-momentum
tensor and the particle current, respectively, measured by a local
observer comoving with a fluid element.  The local values of the
Lorentz-invariant thermodynamic parameters $\beta(x)$ and $\alpha(x)$
can now be fixed by requiring that the average values of the
operators $\hat{\epsilon}{(x)}$ and $\hat{n}{(x)}$ match the local-equilibrium values of these quantities. 
{These so-called {\it {Landau matching conditions}}}~\cite{zubarev1974nonequilibrium,zubarev1997statistical}
are then written as
\bea\label{eq:matching}
\langle \hat{\epsilon}(x)\rangle
=\langle\hat{\epsilon}(x)\rangle_l,\qquad
\langle\hat{n}(x)\rangle
=\langle\hat{n}(x)\rangle_l,
\eea
where for an arbitrary operator $\hat{X}(x)$ the non-equilibrium and local-equilibrium 
statistical averages are defined as
\bea\label{eq:stat_av}
\langle\hat{X}(x)\rangle =
\Tr[\hat{\rho}(t)\hat{X}(x)],\qquad
\langle\hat{X}(x)\rangle_l =
\Tr\big[\hat{\rho}_l(t)\hat{X}(x)\big].
\eea

{Finally, the~fluid 4-velocity $u^\mu$ can be determined by relating it} to a particular physical current.
For example, in the Landau--Lifshitz frame, the 4-velocity is
parallel to the fluid 4-momentum or, equivalently, to the energy flow,
i.e.,
$u_\mu\langle\hat{T}^{\mu\nu}\rangle = \langle\hat{\epsilon}\rangle
u^\nu$~\cite{Landau:Hydro}.
In the Eckart frame, the fluid velocity is associated with the particle
flow via
$\langle\hat{N}^\mu\rangle = \langle\hat{n}\rangle
u^\mu$~\cite{1940PhRv...58..919E}.
However, in the following, we  keep the fluid velocity generic
without specifying any particular reference frame.

The next step is to expand the NESO around the local-equilibrium value in Equation
\eqref{eq:stat_op_local} treating the non-equilibrium part, which~is
described by the operator $\hat{B}$, as a perturbation
\bea  \label{eq:stat_full_2nd_order}
\hat{\rho} = \hat{\rho}_l+\hat{\rho}_1+\hat{\rho}_2,
\eea
where the first-order term is given by
\bea\label{eq:rho_1_final}
\hat{\rho}_1 (t) =  \int d^4x_1\int_0^1d\tau 
\left[\hat{C}_\tau(x_1)
 - \langle \hat{C}_\tau(x_1)\rangle_l\right]\hat{\rho}_{l},
\eea
while the second-order term is
\bea\label{eq:rho_2_final}
\hat{\rho}_2 (t) 
= \frac{1}{2}\int d^4x_1d^4x_2
\int_0^1  d\tau \int_0^1 d\lambda 
 \Big[\tilde{T} \{
\hat{C}_\lambda(x_1)
\hat{C}_\tau(x_2)\} -\langle \tilde{T}\{ 
\hat{C}_\lambda(x_1)
\hat{C}_\tau(x_2)\}\rangle_l \nonumber\\
-
 \langle \hat{C}_\lambda(x_1)\rangle_l  \hat{C}_\tau(x_2) -\hat{C}_\lambda(x_1)
\langle \hat{C}_\tau( x_2)\rangle_l  
+ 2 \langle \hat{C}_\lambda( x_1)\rangle_l  \langle \hat{C}_\tau(x_2)\rangle_l\Big]\hat{\rho}_{l}.
\eea
Here, $\tilde{T}$ is the anti-chronological operator acting on the variables 
$\tau$ and $\lambda$ and we used the short-hand notations
\bea\label{eq:int_short}
\int d^4x_1 = \int d^3x_1\int_{-\infty}^tdt_1 e^{\varepsilon(t_1-t)},
\qquad 
\hat{X}_\alpha = e^{-\alpha {\hat{A}}} \hat{X}e^{\alpha {\hat{A}}},
\qquad 
\alpha \in \tau, \lambda.
\eea

The expansion of Equation \eqref{eq:stat_full_2nd_order} implies that 
the statistical average of any operator 
$\hat{X}(x)$ can be decomposed into three terms 
\bea\label{eq:stat_average}
\langle \hat{X}(x)\rangle 
= \langle \hat{X}(x)\rangle_l +
\langle \hat{X}(x)\rangle_1 +
\langle \hat{X}(x)\rangle_2,
\eea
where the first-order term is given by
\bea\label{eq:stat_average_1}
\langle \hat{X}(x)\rangle_1=
\int d^4x_1
\Big(\hat{X}(x),\hat{C}(x_1)\Big),
\eea
with 
\bea\label{eq:2_point_corr}
\Big(\hat{X}(x),\hat{Y}(x_1)\Big) =
\int_0^1 d\tau \langle\hat{X}(x)
\left[\hat{Y}_\tau(x_1)
 - \langle \hat{Y}_\tau(x_1)\rangle_l\right]\rangle_l
\eea
being the two-point correlation function between two arbitrary
operators~\cite{1984AnPhy.154..229H,2011AnPhy.326.3075H}.
The second-order term in Equation~\eqref{eq:stat_average} can be written as
\bea\label{eq:stat_average_2}
\langle \hat{X}(x)\rangle_2
=\int d^4x_1d^4x_2
\Big(\hat{X}(x),\hat{C}(x_1),\hat{C}(x_2)\Big),
\eea
where we introduced the three-point correlation function 
of the operators $\hat{X}$, $\hat{Y}$, and $\hat{Z}$ as
\bea\label{eq:3_point_corr}
\Big(\hat{X}(x),\hat{Y}(x_1),\hat{Z}(x_2)\Big) &=&
\frac{1}{2}
\int_0^1  d\tau \int_0^1 d\lambda 
\langle \tilde{T} \hat{X}(x)\Big[\hat{Y}_\lambda(x_1)\hat{Z}_\tau(x_2)- \langle \tilde{T}\hat{Y}_\lambda(x_1)\hat{Z}_\tau(x_2)\rangle_l \nonumber\\
&-&
 \langle \hat{Y}_\lambda(x_1)\rangle_l  
\hat{Z}_\tau(x_2) - \hat{Y}_\lambda( x_1) 
\langle\hat{Z}_\tau(x_2)\rangle_l 
+ 2\langle \hat{Y}_\lambda( x_1)\rangle_l  \langle \hat{Z}_\tau(x_2)\rangle_l\Big]\rangle_l.\qquad
\eea

\section{Relativistic Fluid Dynamics at First Order {in Gradients}}
\label{sec:1_order}
 
To examine specific dissipative processes, i.e.,
 {heat conduction}, {particle diffusion, and}
shear and bulk stresses, the~energy-momentum tensor and the
particle current are decomposed as
\bea \label{eq:T_munu_decomp}
\hat{T}^{\mu\nu} &=& \hat{\epsilon} u^{\mu}u^{\nu} - \hat{p}\Delta^{\mu\nu} + \hat{q}^{\mu}u^{\nu}+ \hat{q}^{\nu}u^{\mu} + \hat{\pi}^{\mu\nu},\\
\label{eq:N_decomp}
\hat{N}^{\mu} &=& \hat{n}u^\mu +\hat{j}^{\mu},
\eea
where the fluid velocity $u_\mu$ is normalized as
$u_\mu u^\mu=1$, and $\Delta^{\mu\nu}=g^{\mu\nu}-u^\mu u^\nu$ is the
projection operator onto the 3-space orthogonal to $u_\mu$.
The energy-momentum tensor in Equation~\eqref{eq:T_munu_decomp} is assumed to be symmetric with respect to its indices.
We remind that $u^\mu$ and $\Delta^{\mu\nu}$ appearing in these expressions are classical fields, i.e., $c$-numbers, 
whereas the rest of the quantities are (microscopic) quantum operators, the~physical identification of which is as follows: 
$\hat{\epsilon}$ is the operator of the energy density; $\hat{n}$ is the operator of particle (number) density; 
$\hat{p}$ is the operator of the pressure; and the dissipative
terms $\hat{\pi}^{\mu\nu}$, $\hat{q}^\mu$, and $\hat{j}^\mu$
 are the shear-stress tensor, the~energy-diffusion flux,
and the particle-diffusion flux,~respectively.

{The tensor} decompositions in Equations~\eqref{eq:T_munu_decomp} and \eqref{eq:N_decomp} have the most 
general form that can be constructed from the fluid velocity and the tensor
$\Delta^{\mu\nu}$. 
The operators of the physical quantities 
on the right-hand sides of Equations~\eqref{eq:T_munu_decomp}
and \eqref{eq:N_decomp} can be written as certain projections 
of the energy-{momentum} tensor and the particle current,
\bea\label{eq:proj1_op}
\hat{\epsilon} = u_\mu u_\nu \hat{T}^{\mu\nu},\qquad
\hat{n} = u_\mu\hat{N}^{\mu},\qquad
\hat{p}=-\frac{1}{3}\Delta_{\mu\nu}
\hat{T}^{\mu\nu},\\
\label{eq:proj2_op}
\hat{\pi}^{\mu\nu} = \Delta_{\alpha\beta}^{\mu\nu} \hat{T}^{\alpha\beta},\qquad
\hat{q}^\mu  = u_\alpha\Delta_{\beta}^{\mu}\hat{T}^{\alpha\beta},\qquad
\hat{j}^{\nu}=\Delta_{\mu}^{\nu} \hat{N}^{\mu},
\hspace{-0.1cm}
\eea
where 
\bea\label{eq:projector_delta4} \Delta_{\mu\nu\rho\sigma}=
\frac{1}{2}\left(\Delta_{\mu\rho}\Delta_{\nu\sigma}
  +\Delta_{\mu\sigma}\Delta_{\nu\rho}\right)
-\frac{1}{3}\Delta_{\mu\nu}\Delta_{\rho\sigma}
\eea
 is a traceless rank{--}four
  projector orthogonal to $u^\mu$.
 The dissipative quantities satisfy the
conditions
\bea \label{eq:orthogonality}
\quad
u_{\nu}\hat{q}^{\nu} = 0,\qquad
u_{\nu}\hat{j}^{\nu} = 0,\qquad 
u_{\nu}\hat{\pi}^{\mu\nu} = 0,\qquad 
\hat{\pi}_{\mu}^\mu=0.
\eea

 In local equilibrium, the averages of these operators vanish~\cite{1979TMP....40..821Z}:
\bea\label{eq:diss_currents_av_eq}
\langle \hat{q}^{\mu}\rangle_l =0,\qquad
\langle \hat{j}^{\mu}\rangle_l =0,\qquad
\langle \hat{\pi}^{\mu\nu}\rangle_l =0,
\eea
and one recovers the limit of ideal {fluid} dynamics. The
local-equilibrium pressure is given by the equation of state,
i.e.,  $\langle\hat{p}\rangle_l\equiv p=p(\epsilon,n)$, which~closes
the set of ideal {fluid-dynamical} equations {of~motion}.

Consider next fluid dynamics at first order {in gradients}. Quite generally,
the {fluid-dynamical} quantities ${\pi}^{\mu\nu}$, ${q}^\mu$, and ${j}^\mu$
are obtained as the statistical averages of the corresponding
operators over the NESO according to
Equations~\eqref{eq:stat_average}--\eqref{eq:3_point_corr}.  Keeping only
the first-order terms in Equation~\eqref{eq:stat_average}, we obtain the
relativistic Navier--Stokes equations
\bea \label{eq:NS_eqs}
{\pi}_{\mu\nu} =
 2\eta \sigma_{\mu\nu},\qquad
\Pi =-\zeta\theta,\qquad
\mathscr{J}_\mu=\kappa\left(\frac{ n T}{h}\right)^2 \nabla_\mu \alpha,
\eea
where ${\pi}_{\mu\nu}\equiv \langle\hat{\pi}_{\mu\nu}\rangle$, 
the bulk viscous pressure
$\Pi\equiv \langle\hat{p}\rangle-\langle\hat{p}\rangle_l$ is the difference
between the first-order average of the pressure operator and the {local-equilibrium}
value of pressure, $h=\epsilon+p$ is the enthalpy density, and
\bea\label{eq:op_heat_flux}
\mathscr{{J}}_{\mu}={j}_{\mu}-\frac{n}{h}{q}_{\mu}
\eea
is the irreversible particle flow, i.e., the~particle flow with
respect to the energy 
flow~\cite{1976AnPhy.100..310I,1979AnPhy.118..341I}. On~the right-hand
sides of Equation~\eqref{eq:NS_eqs},
$\sigma_{\mu\nu}=\partial_{<\alpha} u_{\beta >}$ is the 
{shear} tensor, where~{angular brackets denote the projection with the projector in Equation
(\ref{eq:projector_delta4}), i.e.,}
$A_{<\mu\nu>}=\Delta_{\mu\nu}^{\alpha\beta}A_{\alpha\beta}$,
$\theta =\partial_\mu u^\mu$ is the {expansion scalar, and}~\mbox{$\nabla_\alpha= \Delta_{\alpha\beta}\partial^\beta$} is the covariant
spatial derivative. The coefficients $\eta$, $\zeta$, and $\kappa$ are
the transport coefficients of the shear and bulk
viscosities, and the thermal conductivity, respectively.
These~transport coefficients can be expressed through two-point
correlation functions via the following Kubo
formulae~\cite{1984AnPhy.154..229H,2011AnPhy.326.3075H,Hayata2015}
\bea \label{eq:shear_def}
\eta &=& \frac{\beta}{10}\int d^4x_1
\Big(\hat{\pi}_{\mu\nu}(x),\hat{\pi}^{\mu\nu}(x_1)\Big),\\
 \label{eq:bulk_def}
\zeta &=& \beta\int d^4x_1
\Big(\hat{p}^*(x), \hat{p}^*(x_1)\Big),\\
 \label{eq:kappa_def}
\kappa &=& -\frac{\beta^2}{3}\int d^4x_1
\left({\hat{h}}_{\mu}
(x),{\hat{h}}^{\mu}(x_1)\right),
\eea 
where
\bea\label{eq:p_star}
\hat{p}^* 
= \hat{p} - \gamma\hat{\epsilon}-\delta\hat{n},\qquad
\hat{h}^\mu= \hat{q}^{\mu}
-\frac{h}{n}\hat{j}^{\mu},
\eea
and
\bea\label{eq:gamma_delta}
\gamma =\left(\frac{\partial p}{\partial\epsilon}
\right)_{n},\qquad
\delta = \left(\frac{\partial p}
{\partial n} \right)_{\epsilon}.
\eea

The correlation functions in
Equations~\eqref{eq:shear_def}--\eqref{eq:kappa_def} {are evaluated
in a uniform background}, i.e., as if the system was in {\it {global thermodynamical
  equilibrium}}. They can be expressed in terms of the two-point
retarded {\it equilibrium Green functions} as~\cite{1984AnPhy.154..229H,2011AnPhy.326.3075H}
\bea \label{eq:shear_kubo}
\eta &=& -\frac{1}{10}\frac{d}{d\omega} {\rm Im}G^R_{\hat{\pi}_{\mu\nu}\hat{\pi}^{\mu\nu}}(\omega)
\bigg\vert_{\omega=0},\\
\label{eq:bulk_kubo}
\zeta &=& -\frac{d}{d\omega} {\rm Im}G^R_{\hat{p}^{*}
\hat{p}^*}(\omega)\bigg\vert_{\omega=0},\\
\label{eq:kappa_kubo}
\kappa & =& \frac{1}{3T}\frac{d}{d\omega} {\rm Im}G^R_{\hat{h}_{\mu}\hat{h}^{\mu}}(\omega)\bigg\vert_{\omega=0},
\eea
where, for any two operators $\hat X$ and $\hat Y$,
\bea\label{eq:green_func}
G^R_{\hat{X}\hat{Y}}(\omega) 
\equiv -i\int_{0}^{\infty} dt
e^{i\omega t}\int d^3x\langle\big[\hat{X}(\bm x, t),
\hat{Y}(\bm 0,0)\big]\rangle_l .
\eea
Equations \eqref{eq:shear_kubo}--\eqref{eq:kappa_kubo} represent a
particularly suitable form {for the} Kubo formulae, which~lends itself to
evaluation {using the} methods of equilibrium finite-temperature field
theory.  

Before closing this section, it is useful to clarify the
relation between the expansions in powers of the thermodynamic forces
and {in powers of} the Knudsen number $K=l/L$, where~$l$ and $L$ are typical
microscopic and macroscopic length scales{, respectively.} To obtain the
relations~in Equation \eqref{eq:NS_eqs} from Equation~\eqref{eq:stat_average_1}, we used
Curie's theorem. It states that, in an isotropic medium, the correlations
between operators of different rank vanish~\cite{de1969non}.
The integrands in Equations~\eqref{eq:shear_def}--\eqref{eq:kappa_def} are
mainly concentrated within the range $|x_1-x|\lesssim l$, where~$l$ is
the mean correlation length, which~in the weak-coupling limit is of
the order of the particle mean free path.  {The fluid-dynamical} regime implies
$l\ll L$, where~$L$ is the typical length scale over which the
parameters $\beta^\nu$ and $\alpha$ vary in space.  Therefore, the~thermodynamic forces $\partial^\mu \beta^\nu$ and $\partial^\mu\alpha$
involved in Equation~\eqref{eq:C_op} can be factored out from the
integral~in Equation \eqref{eq:stat_average_1} with their average values at $x$,
i.e., the~{\it non-locality} of the thermodynamic forces can be
neglected in this approximation.  Because
$|\sigma^{\rho\sigma}|\simeq |u^{\rho}|/L$,
the relations~ in Equation \eqref{eq:NS_eqs} obtained from the gradient
expansion~in Equation \eqref{eq:stat_full_2nd_order} of the NESO are consistent
with the expansion scheme in powers of the Knudsen~number.

\section{Relativistic Fluid Dynamics at Second Order {in Gradients}}
\label{sec:2_order}

We have computed systematically all second-order corrections to the
dissipative quantities ${\pi}^{\mu\nu}$, ${\Pi}$, and
${\mathscr J}^\mu$ based on
Equations~\eqref{eq:stat_average}--\eqref{eq:3_point_corr}~\cite{Hydro_preprint,
  ArusPhD}. Before presenting the results, we note that the
second-order contributions arise not only from
Equation~\eqref{eq:stat_average_2}, which~is quadratic in the thermodynamic
force $\hat{C}$, but also from Equation~\eqref{eq:stat_average_1}, where
the non-local nature of the thermodynamic forces in space and time
should be carefully taken into account. The non-local effects generate
finite relaxation terms in the {fluid-dynamical} equations, which~are
required for causality.  To see that these corrections are of 
second order in the Knudsen number, note that they involve the
differences of the thermodynamic forces, {e.g.,}
$\partial^{\mu}\beta^{\nu}$, at the points $x_1$ and $x$  {(}see
Equations~\eqref{eq:C_op} and \eqref{eq:stat_average_1}{)}.  Therefore, we
can approximate 
$\partial^{\mu}\beta^{\nu}(x_1)-\partial^{\mu}\beta^{\nu}(x)\simeq \partial_\lambda\partial^{\mu}\beta^{\nu}(x)(x_1-x)^\lambda\sim
K\partial^{\mu}\beta^{\nu}(x)$,
because $x_1-x\sim l$ and $\partial\sim L^{-1}$, as already done in
Section~\ref{sec:1_order}. Thus, these corrections contain an additional
power of the Knudsen number $K$ as compared to the first-order
expressions~in Equation \eqref{eq:NS_eqs} and, therefore, are of second~order.

Here, we restrict ourselves to the second-order expression for
the shear-stress tensor and compare it with the results of References~\cite{2008JHEP...04..100B,2010CQGra..27b5006R}.

\subsection{Second-Order Corrections to the Shear-Stress Tensor}
\label{sec:2nd_shear}

As explained above, we now keep the NESO at second order in small
perturbations from local equilibrium and, in addition, we retain terms
which are of second order in the gradients of thermodynamic forces.
In this manner, we find the shear-stress tensor at second order as
\bea\label{eq:shear_total_final}
{\pi}_{\mu\nu}
=  2\eta \sigma_{\mu\nu}
 -2\eta\tau_\pi (\dot{\sigma}_{\mu\nu}+ \gamma \theta \sigma_{\mu\nu})
 + \lambda_\pi \sigma_{\alpha<\mu}\sigma_{\nu>}^{\alpha}
 + 2\lambda_{\pi\Pi}\theta \sigma_{\mu\nu}
 + \lambda_{\pi{\mathscr J}}\nabla_{<\mu}\alpha\nabla_{\nu>}\alpha,
\eea
where
$\dot{\sigma}_{\mu\nu}\equiv\Delta_{\mu\nu\rho\sigma}D
\sigma^{\rho\sigma}$,
with $D=u^\mu\partial_\mu$ being the {comoving} derivative, and
$\tau_\pi$, $\lambda_\pi$, $\lambda_{\pi\Pi}$, and
$\lambda_{\pi{\mathscr J}}$ represent four new coefficients
associated with the second-order corrections to the shear stress.  The~first term on the right-hand side of Equation~\eqref{eq:shear_total_final}
is easily recognized as the first-order (Navier--Stokes) contribution.
The second-order terms collected in the parentheses (i.e., those
$\propto\tau_\pi$) represent the non-local corrections to
Equation~\eqref{eq:stat_average_1}, whereas the last three terms stand for
the nonlinear corrections arising from the three-point correlation
functions in Equation~\eqref{eq:stat_average_2}. The
first non-local correction describes {\it memory effects} due to its
non-locality in time.  The relevant transport coefficient $\tau_\pi$,
which has the dimension of time, measures how long the information
remains in the ``memory'' of the shear-stress tensor
$\pi_{\mu\nu}$. Therefore, it is natural to associate it with the
relaxation time of the shear stresses towards their asymptotic
Navier--Stokes values.  The second term involves a product of $\sigma_{\mu\nu}$ 
with $\theta=\partial_\mu u^\mu$ and can be regarded as a (scalar)
measure of the spatial ``non-locality'' in the fluid-velocity
field. This term describes how the shear-stress tensor is distorted by
uniform expansion or contraction of the fluid.

We find that the relaxation time $\tau_\pi$ is related to the frequency derivative of
the corresponding first-order transport coefficient, i.e., 
the shear viscosity, by a Kubo formula
\bea\label{eq:tau_pi}
\eta\tau_\pi = -i\frac{d}{d\omega}\eta(\omega)\bigg\vert_{\omega=0}
=\frac{1}{10}\frac{d^2}{d\omega^2} {\rm Re}G^R_{\hat{\pi}_{\mu\nu}
\hat{\pi}^{\mu\nu}}(\omega)\bigg\vert_{\omega=0},
\eea
where $\eta\equiv \eta(0)$ is given by Equation~\eqref{eq:shear_kubo}, the
retarded Green's function $G^R_{\hat{\pi}_{\mu\nu}\hat{\pi}^{\mu\nu}}$ is defined in
  Equation~\eqref{eq:green_func}, and the frequency-dependent shear
  viscosity $\eta(\omega)$ is given by Equation~\eqref{eq:shear_kubo} for non-vanishing $\omega$.
  Similar expressions for the relaxation times were obtained previously in
References~\cite{2008JHEP...04..100B,2010CQGra..27b5006R,2011PhRvL.106l2302M,2017PhRvC..95f4906C}.
  
  The physical meaning of the Equation~\eqref{eq:tau_pi} for $\tau_\pi$
  is easy to understand.  As {mentioned above}, the~relaxation terms
  originate from the non-local (memory) effects encoded in the
  non-equilibrium statistical operator. In the case where these memory
  effects are neglected (first-order theory), the~proportionality
  between $\pi_{\mu\nu}$ and $\sigma_{\mu\nu}$ is given by the
  zero-frequency (static) limit of the shear viscosity, as seen from
  Equations~\eqref{eq:NS_eqs} and \eqref{eq:shear_kubo}. The memory effects
  imply a time delay, which~translates into a frequency dependence of the
  shear viscosity \cite{Lublinsky}. At leading order, this is accounted for by the
  first-order frequency derivative of $\eta(\omega)$ as
  Equation~\eqref{eq:tau_pi} demonstrates.

  The last three terms in Equation~\eqref{eq:shear_total_final} contain all
  combinations of the thermodynamic forces $\sigma_{\mu\nu}$,
  $\theta$, and $\nabla_\mu\alpha$ which are allowed by the
  symmetries to quadratic order. These are $\theta {\sigma}_{\mu\nu}$,
  $\sigma_{\rho<\mu}\sigma_{\nu>}^{\rho}$, and~$\nabla_{<\mu}\alpha\nabla_{\nu>}\alpha$.  The second-order
  transport coefficients associated with these terms can be
  expressed via three-point correlation functions according to
\bea\label{eq:lambda_pi}
\lambda_\pi &=& \frac{12}{35}\beta^2\int d^4x_1d^4x_2
\Big(\hat{\pi}_{\mu}^{\nu}(x),\hat{\pi}_{\nu}^{\lambda}(x_1),\hat{\pi}_{\lambda}^{\mu}(x_2)\Big),\\
\label{eq:lambda_2}
\lambda_{\pi\Pi} &=& -\frac{\beta^2 }{5}\int d^4x_1d^4x_2
\Big(\hat{\pi}_{\mu\nu}(x),\hat{\pi}^{\mu\nu}(x_1),\hat{p}^*(x_2)\Big),\\
\label{eq:lambda1_ab}
\lambda_{\pi{\mathscr J}} &=& \frac{1}{5}\int d^4x_1d^4x_2
\Big(\hat{\pi}_{\mu\nu}(x),
\hat{\mathscr J}^{\mu}(x_1),
\hat{\mathscr J}^{\nu}(x_2)\Big),
\eea
where $\hat{\mathscr J}^{\mu}$ is the operator corresponding to the
4-current~\eqref{eq:op_heat_flux}.  In analogy to the leading-order
coefficient $\eta$, which~is given by the two-point correlation of the
{shear-stress} tensor, the~second-order coefficient $\lambda_\pi$ is given by
the three-point correlation of the shear-stress tensor.  The
coefficient $\lambda_{\pi\Pi}$ describes the nonlinear coupling
between shear- and bulk-viscous processes and is given by a
three-point correlation function between two {shear-stress tensors} and the
bulk viscous pressure.  Finally, the~coefficient
$\lambda_{\pi{\mathscr J}}$ describes the nonlinear coupling between
the shear and the diffusion processes. Similarly, this coefficient is
given by a three-point correlation function between two diffusion
currents and the shear-stress tensor. Note that, in
Equation~\eqref{eq:shear_total_final}, the term $\propto \lambda_{\pi\Pi}$
and the second term in parenthesis have the same gradient
structure, but they have different origins and physical
interpretation. The term $\propto \tau_\pi$ originates from {\it
  non-local} effects in the statistical distribution, whereas the term
$\propto \lambda_{\pi\Pi}$ stands purely for the {\it nonlinear} coupling
between the bulk- and the shear-viscous effects.  In this sense, it is
natural to regard as nonlinear only the term
$\propto \lambda_{\pi\Pi}$, but not the term $\propto\tau_\pi$. A
similar classification of the second-order terms was suggested earlier 
in Reference~\cite{MooreSohrabi2012}.

\subsection{Comparison with Previous Studies}
\label{sec:shear_compare}

For the sake of simplicity we  consider here a fluid without
conserved charges. In this case, Equation~\eqref{eq:gamma_delta}  implies 
$\gamma \equiv c_s^2$, where~$c_s$ is the speed of sound, and 
Equation~\eqref{eq:shear_total_final} reduces to
\bea\label{eq:shear_discuss}
{\pi}_{\mu\nu}
=  2\eta \sigma_{\mu\nu}
 -2\eta\tau_\pi (\dot{\sigma}_{\mu\nu}+ c_s^2 \theta \sigma_{\mu\nu}) + \lambda_\pi \sigma_{\alpha<\mu}\sigma_{\nu>}^{\alpha} + 2\lambda_{\pi\Pi}\theta \sigma_{\mu\nu}.
\eea

\textls[-15]{Baier et al. \cite{2008JHEP...04..100B} {found} in this
  case and for conformal fluids
\bea\label{eq:shear_Baier2008}
  \pi_{\mu\nu}^{B} = 2\eta \sigma_{\mu\nu}
- 2\eta\tau_\pi \left( \dot{\sigma}_{\mu\nu} + \frac{1}{3}\theta\sigma_{\mu\nu} \right) + \lambda_1 {\sigma}_{\alpha <\mu} \sigma_{\nu>}^{\alpha},
  \eea 
  where we have {neglected} terms {proportional to} the vorticity
  tensor
  $w_{\alpha\beta}=(\nabla_{\alpha}u_{\beta}-\nabla_{\beta}u_{\alpha})/2$.
  ({{Note that} Baier et al. \cite{2008JHEP...04..100B} and
    Romatschke~\cite{2010CQGra..27b5006R}} {used} a metric
  convention opposite to ours, and their definition of the shear
  viscosity differs from ours by a factor~of~2}).  Because
$ c_s^2=1/3$ for a conformal fluid, we recover from
Equation~\eqref{eq:shear_discuss} the term {proportional to}
$\tau_\pi$ in Equation~\eqref{eq:shear_Baier2008}.  Furthermore,
because conformal invariance implies a vanishing bulk viscous
pressure, the~correlations involving the operator $\hat{p}^*$ (see
Equations~\eqref{eq:bulk_def} and \eqref{eq:p_star}) vanish, i.e.,
$\lambda_{\pi\Pi}=0$ in this case. Finally we see that
$\lambda_1\equiv\lambda_{\pi}$.

In the case of non-conformal fluids, the second-order expression for
the shear-stress tensor was found, {e.g.,} in Reference~\cite{2010CQGra..27b5006R}
in the absence of conserved charges. Again, neglecting the vorticity
tensor and assuming flat space-time, 
\bea\label{eq:shear_Romatschke2010}
  \pi_{\mu\nu}^R = 2\eta \sigma_{\mu\nu}
- 2\eta\tau_\pi \left( \dot{\sigma}_{\mu\nu} + \frac{1}{3}\theta\sigma_{\mu\nu} \right) + \lambda_1 {\sigma}_{\alpha <\mu} \sigma_{\nu>}^{\alpha}
    -\frac{2}{3}\eta\tau_\pi^*\theta\sigma_{\mu\nu}+
  \lambda_4\nabla_{<\mu}\ln s\nabla_{\nu>}\ln s.
\eea 
The term $\propto\tau_\pi^*$ has the same gradient structure as the
non-local term $-2\eta\tau_\pi\theta\sigma_{\mu\nu}/3$. Comparing~Equation~\eqref{eq:shear_Romatschke2010} with our
expression~in Equation \eqref{eq:shear_discuss}, we identify
$\tau^*_\pi=\tau_\pi(3c_s^2-1)-3\lambda_{\pi\Pi}/\eta$, and~$\lambda_4=0$.

We also note that Equation~\eqref{eq:shear_total_final} does not contain
{terms proportional to the vorticity}. To derive {such} terms,
one needs to include {an initial} non-zero angular momentum in the
local-equilibrium distribution~\cite{2017JHEP...10..091B}.

\subsection{Relaxation Equation for the Shear-Stress Tensor}

A relaxation-type equation for ${\pi}_{\mu\nu}$ can now be derived
from Equation~\eqref{eq:shear_total_final}. For~this purpose, we replace the first-order expression
$2\sigma^{\rho\sigma}\to \eta^{-1}{\pi}^{\rho\sigma}$ in the second
term on the right-hand-side of Equation~\eqref{eq:shear_total_final}, as
has also been done in
References~\cite{2008JHEP...04..100B,2013PhRvC..87e1901J,2015JHEP...02..051F}.
This substitution is justified {up to} second order in space-time gradients.
We then obtain
\bea\label{eq:shear_relax}
-2\eta\tau_\pi \dot{\sigma}_{\mu\nu}
\simeq  -\tau_\pi \dot{\pi}_{\mu\nu}+ \tau_\pi\beta\eta^{-1} \left(\gamma \frac{\partial\eta}{\partial\beta} -\delta \frac{\partial\eta}{\partial\alpha}\right)\theta {\pi}_{\mu\nu}\simeq
-\tau_\pi \dot{\pi}_{\mu\nu}+ 2\tau_\pi\beta \left(\gamma \frac{\partial\eta}{\partial\beta} -\delta \frac{\partial\eta}{\partial\alpha}\right)\theta {\sigma}_{\mu\nu},
\eea
where $\dot{\pi}_{\mu\nu}=\Delta_{\mu\nu\rho\sigma}D{\pi}^{\rho\sigma}$.
The terms in brackets contain the corresponding partial derivatives of
$\eta$, which~in general are not small and should not be neglected.
In Equation~\eqref{eq:shear_relax}, we employed the relations
$D\beta=\beta\theta \gamma$ and
$D\alpha =-\beta\theta\delta$~\cite{2011AnPhy.326.3075H}.  Combining
Equations~\eqref{eq:shear_total_final} and \eqref{eq:shear_relax} and
introducing the~coefficient
\bea\label{eq:lambda}
\lambda = \lambda_{\pi\Pi}-\gamma\eta\tau_\pi
+\tau_\pi \beta \left(\gamma \frac{\partial\eta}{\partial\beta} -\delta \frac{\partial\eta}{\partial\alpha}\right),
\eea
we finally obtain 
\bea\label{eq:shear_final}
\tau_\pi \dot{\pi}_{\mu\nu}+ {\pi}_{\mu\nu}
= 2\eta \sigma_{\mu\nu}
 + 2\lambda \theta {\sigma}_{\mu\nu} + \lambda_\pi \sigma_{\rho<\mu}\sigma_{\nu>}^{\rho} +\lambda_{\pi \mathscr J}\nabla_{<\mu}\alpha\nabla_{\nu>}\alpha.
\eea

The time-derivative term on the left-hand side describes the
relaxation of the shear-stress tensor towards its Navier--Stokes value
on the characteristic time scale $\tau_\pi$. Indeed, for vanishing
right-hand side the relaxation is exponential, $\pi_{\mu\nu}\propto
\exp(-t/\tau_\pi)$, with a characteristic relaxation time scale
$\tau_\pi$.  We~would like to stress that the exponential relaxation
over a time scale $\tau_\pi$ is a direct consequence of the
substitution $2\sigma^{\rho\sigma}\to \eta^{-1}{\pi}^{\rho\sigma}$
made above; it is not a manifestation of a specific relaxation process
on the microscopic level.  However, a direct way to obtain such a
relaxation term is via the method of moments applied to the Boltzmann
equation \cite{Denicol2012}. In this case, the~time scale $\tau_\pi$
is an intrinsic property of the collision kernel in the Boltzmann
equation.

\section{Summary}
\label{sec:summary}

This work concisely presents the derivation of second-order
relativistic {dissipative} fluid dynamics within Zubarev's NESO
formalism -- a
method which is well-suited for treatments of strongly correlated
systems. The simple case of a one-component fluid without
electromagnetic fields or vorticity in flat space-time was considered here.

Our analysis shows that the {second-order 
dissipative terms} arise from: (i) the quadratic terms in the Taylor
expansion of the statistical operator; and (ii) the linear terms of the
same expansion which include memory and non-locality in space. In
particular, we find that the type-(ii) terms describe the relaxation
in time of the dissipative fluxes, which~is essential for the
causality of the {fluid-dynamical}~theory. 
 
Using the NESO method and the example of the {shear-stress tensor}, we
demonstrated that the second-order transport coefficients can be
expressed in terms of certain two- and three-point equilibrium
correlation functions. A discussion of the transport coefficients
associated with other thermodynamic fluxes can be found
elsewhere~\cite{ArusPhD}. Furthermore, we have shown that Kubo-type
formulae for the relaxation times of the dissipative fluxes can be
obtained within the NESO formalism  {(}see Equation~\eqref{eq:tau_pi}{)}.  These
are given by the zero-frequency limit of the derivatives of the
corresponding first-order transport coefficients with respect to the
frequency. These can be computed from {the theory of
quantum fields in equilibrium} at non-zero temperature as, for example, was done by us
for the QGP within the
Nambu--Jona--Lasinio model~\cite{Harutyunyan2017a,Harutyunyan2017b}.

\vspace{6pt}

\authorcontributions{Conceptualization, A. H., A. S., D. R.;
Methodology, A. H., A. S., D. R.;
Investigation, A. H., A. S., D. R.;
Writing—Original Draft Preparation, A. H., A. S., D. R.;
Writing—Review \& Editing, A. H., A. S., D. R;
Funding Acquisition, A. S., D. R.}


\acknowledgments{We are grateful to {Ulrich Heinz and Xu-Guang Huang}
for discussions.~A.H. acknowledges support from the HGS-HIRe graduate program at  Goethe University.~A.S.~was supported by the Deutsche~Forschungsgemeinschaft (Grant No.~SE~1836/4-1).~D.H.R. acknowledges support by the High-End Visiting Expert project GDW20167100136 of the State Administration of Foreign Experts Affairs (SAFEA) of China and by the   Deutsche Forschungsgemeinschaft (DFG) through the grant  CRC-TR 211 ``Strong-interaction matter under extreme~conditions''. }

\conflictsofinterest{{The authors declare no conflict of interest.}}

\reftitle{References}


\begin{thebibliography}{999}
\providecommand{\natexlab}[1]{#1}

\bibitem[Landau and Lifshitz(1987)]{Landau:Hydro}
Landau, L.; Lifshitz, E.
\newblock {\em Fluid Mechanics}; Butterworth-Heinemann: Oxford, UK, 1987.

\bibitem[{Denicol}(2014)]{2014JPhG...41l4004D}
{Denicol}, G.S.
\newblock {Kinetic foundations of relativistic dissipative fluid dynamics}.
\newblock {\em J. Phys. G Nucl. Phys.} {\bf 2014}, 
{\em 41},~124004.


\bibitem{Denicol2012} 
  Denicol, G.S.; Niemi, H.; Molnar, E.; Rischke, D.H. 
  \newblock {Derivation of transient relativistic fluid dynamics from the Boltzmann equation}. 
  \newblock {\em Phys.\ Rev.\ D} {\bf 2012}, {\em 85}, 039902. 

  
\bibitem[{Florkowski} \em{et~al.}(2017){Florkowski}, {Heller}, and
  {Spalinski}]{2017arXiv170702282F}
{Florkowski}, W.; {Heller}, M.P.; {Spalinski}, M.
\newblock {New theories of relativistic hydrodynamics in the LHC era}.
\newblock \emph{Rep.~Prog.~Phys.} \textbf{2018}, \emph{81}, 046001.


\bibitem[{Romatschke}(2010)]{2010IJMPE..19....1R}
{Romatschke}, P.
\newblock {New Developments in Relativistic Viscous Hydrodynamics}.
\newblock {\em Int. J. Mod. Phys. E} {\bf 2010}, {\em
  19},~1--53.


\bibitem[Zubarev(1974)]{zubarev1974nonequilibrium}
Zubarev, D.
\newblock {\em Nonequilibrium Statistical Thermodynamics}; Studies in Soviet
  Science; Consultants Bureau: {London,~UK}, 1974.


\bibitem[Zubarev \em{et~al.}(1997)Zubarev, Morozov, and
  R{\"o}pke]{zubarev1997statistical}
Zubarev, D.; Morozov, V.; R{\"o}pke, G.
\newblock {\em Statistical Mechanics of Nonequilibrium Processes}; John Wiley
  \& Sons: {Hoboken, NJ, USA}, 1997.


\bibitem[{Hosoya} \em{et~al.}(1984){Hosoya}, {Sakagami}, and
  {Takao}]{1984AnPhy.154..229H}
{Hosoya}, A.; {Sakagami}, M.A.; {Takao}, M.
\newblock {Non\-equilibrium thermodynamics in field theory: Transport
  coefficients}.
\newblock {\em Ann. Phys.} {\bf 1984}, {\em 154},~229--252.

\bibitem[{Huang} \em{et~al.}(2011){Huang}, {Sedrakian}, and
  {Rischke}]{2011AnPhy.326.3075H}
{Huang}, X.G.; {Sedrakian}, A.; {Rischke}, D.H.
\newblock {Kubo formulas for relativistic fluids in strong magnetic fields}.
\newblock {\em Ann. Phys.} {\bf 2011}, {\em 326},~3075--3094.


\bibitem[{Hayata} \em{et~al.}(2015){Hayata}, {Hidaka}, {Noumi}, and                                            
  {Hongo}]{Hayata2015}                                                                                         
{Hayata, T.; Hidaka, Y.; Noumi, T.; Hongo, M.  }                                                         
\newblock {Relativistic hydrodynamics from quantum field theory on the basis of
  the generalized Gibbs ensemble method}.
\newblock {\em Phys. Rev. D} {\bf 2015}, {\em 92},~065008.


\bibitem[{Becattini} \em{et~al.}(2015){Becattini}, {Bucciantini}, {Grossi}, and                                
  {Tinti}]{Becattini2015}                                                                                      
{Becattini}, F.; {Bucciantini}, L.; {Grossi}, E.; {Tinti}, L.                                                  
\newblock {Local thermodynamical equilibrium and the frame for a quantum
  relativistic fluid}.
\newblock {\em Eur. Phys. J. C} {\bf 2015}, {\em 75},~191.

\bibitem[{Hongo}(2017)]{Hongo2017}
{Hongo}, M.
\newblock {Path-integral formula for local thermal equilibrium}.
\newblock {\em Ann. Phys.} {\bf 2017}, {\em 383},~1--32.

\bibitem[{Harutyunyan} \em{et~al.}(){Harutyunyan}, {Sedrakian}, and
  {Rischke}]{Hydro_preprint}
{Harutyunyan}, A.; {Sedrakian}, A.; {Rischke}, D.H.~{Second-order relativistic hydrodynamics from a non-equilibrium
  statistical operator}. 
\newblock {\em \rm In preparation\hspace{-.12cm}}.

\bibitem[Harutyunyan(2017)]{ArusPhD}
Harutyunyan, A.
\newblock Relativistic Hydrodynamics and Transport in Strongly Correlated
  Systems.
\newblock Ph.D. Thesis, Goethe University, Frankfurt am Main, Germany,  2017.

\bibitem[{Eckart}(1940)]{1940PhRv...58..919E}
{Eckart}, C.
\newblock {The Thermodynamics of Irreversible Processes. III. Relativistic
  theory of the simple fluid}.
\newblock {\em Phys.~Rev.} {\bf 1940}, {\em 58},~919--924.

\bibitem[{Israel}(1976)]{1976AnPhy.100..310I}
{Israel}, W.
\newblock {Nonstationary irreversible thermodynamics: A causal relativistic
  theory}.
\newblock {\em Ann. Phys.} {\bf 1976}, {\em 100},~310--331.

\bibitem[{Israel} and {Stewart}(1979)]{1979AnPhy.118..341I}
{Israel}, W.; {Stewart}, J.M.
\newblock {Transient relativistic thermodynamics and kinetic theory}.
\newblock {\em Ann. Phys.} {\bf 1979}, {\em 118},~341--372.

\bibitem[{Baier} \em{et~al.}(2008){Baier}, {Romatschke}, {Thanh Son},
  {Starinets}, and {Stephanov}]{2008JHEP...04..100B}
{Baier}, R.; {Romatschke}, P.; {Thanh Son}, D.; {Starinets}, A.O.; {Stephanov},
  M.A.
\newblock {Relativistic viscous hydrodynamics, conformal invariance, and
  holography}.
\newblock {\em J. High Energy Phys.} {\bf 2008}, {\em 4},~100.

\bibitem[{Moore} and {Sohrabi}(2012)]{MooreSohrabi2012}
{Moore}, G.D.; {Sohrabi}, K.A.
\newblock {Thermodynamical second-order hydrodynamic coefficients}.
\newblock {\em J. High Energy Phys.} {\bf 2012}, {\em 11},~148.

\bibitem[{Zubarev} \em{et~al.}(1979){Zubarev}, {Prozorkevich}, and
  {Smolyanskii}]{1979TMP....40..821Z}
{Zubarev}, D.N.; {Prozorkevich}, A.V.; {Smolyanskii}, S.A.
\newblock {Derivation of nonlinear generalized equations of quantum
  relativistic hydrodynamics}.
\newblock {\em Theor. Math. Phys.} {\bf 1979}, {\em
  40},~821--831.


\bibitem[de~Groot and Mazur(1969)]{de1969non}
De~Groot, S.; Mazur, P.~{\em Non-Equilibrium Thermodynamics}; 
{Interscience Publishers: Amsterdam, The~Netherlands}, 
1969.


\bibitem[{Romatschke}(2010)]{2010CQGra..27b5006R}
{Romatschke}, P.
\newblock {Relativistic viscous fluid dynamics and non-equilibrium entropy}.
\newblock {\em Class. Quantum Gravity} {\bf 2010}, {\em 27},~025006.

  
\bibitem[{Moore} and {Sohrabi}(2011)]{2011PhRvL.106l2302M}
{Moore}, G.D.; {Sohrabi}, K.A.
\newblock {Kubo Formulas for Second-Order Hydrodynamic Coefficients}.
\newblock {\em Phys. Rev. Lett.} {\bf 2011}, {\em 106},~122302.


\bibitem[{Czajka} and {Jeon}(2017)]{2017PhRvC..95f4906C}
{Czajka}, A.; {Jeon}, S.
\newblock {Kubo formulas for the shear and bulk viscosity relaxation times and
  the scalar field theory shear {$\tau$}$_{\pi}$ calculation}.
\newblock {\em Phys. Rev. D} {\bf 2017}, {\em 95},~064906.

  
\bibitem{Lublinsky} 
 Lublinsky, M.; Shuryak, E.
{ Improved Hydrodynamics from the AdS/CFT}.
  {\em Phys.\ Rev.\ D} {\bf 2009},  {\em 80}, 065026.


\bibitem[{Buzzegoli}, {Grossi} and {Becattini}(2017)]{2017JHEP...10..091B}
{Buzzegoli}, M.; {Grossi}, E.; {Becattini}, F.
\newblock {General equilibrium second-order hydrodynamic coefficients for free quantum fields}.
\newblock {\em J. High Energy Phys.} {\bf 2017}, {\em 10},~091.


\bibitem[{Jaiswal}(2013)]{2013PhRvC..87e1901J}
{Jaiswal}, A.
\newblock {Relativistic dissipative hydrodynamics from kinetic theory with
  relaxation-time approximation}.
\newblock {\em Phys.\ Rev.\ D} {\bf 2013}, {\em 87},~051901.


\bibitem[{Finazzo} \em{et~al.}(2015){Finazzo}, {Rougemont}, {Marrochio}, and
  {Noronha}]{2015JHEP...02..051F}
{Finazzo}, S.I.; {Rougemont}, R.; {Marrochio}, H.; {Noronha}, J.
\newblock {Hydrodynamic transport coefficients for the non-conformal
  quark-gluon plasma from holography}.
\newblock {\em J. High Energy Phys.} {\bf 2015}, {\em 2},~51.


\bibitem[{Harutyunyan} \em{et~al.}(2017){Harutyunyan}, {Rischke}, and
  {Sedrakian}]{Harutyunyan2017a}
{Harutyunyan}, A.; {Rischke}, D.H.; {Sedrakian}, A.
\newblock {Transport coefficients of two-flavor quark matter from the Kubo
  formalism}.
\newblock {\em \prd} {\bf 2017}, {\em 95},~114021.


\bibitem[{Harutyunyan} and {Sedrakian}(2017)]{Harutyunyan2017b}
{Harutyunyan}, A.; {Sedrakian}, A.
\newblock {Bulk viscosity of two-flavor quark matter from the Kubo formalism}.
\newblock {\em \prd} {\bf 2017}, {\em 96},~034006.


\end{thebibliography}
\end{document}